\documentclass[twocolumn,twoside,english,aps,prb,groupedaddress,amsmath,amssymb,floatfix]{revtex4-1}

\usepackage{graphicx}
\usepackage{float}
\usepackage{physics}
\usepackage{amsmath}
\usepackage{dcolumn}
\usepackage{bm}
\usepackage{units}
\usepackage[version=4]{mhchem}
\usepackage{color}
\usepackage{array}
\usepackage[dvipsnames]{xcolor}
\usepackage{soul}
\usepackage{listings}
\usepackage{url}
\usepackage[colorlinks = true, linkcolor = blue, urlcolor  = blue, citecolor = blue, anchorcolor = blue, unicode]{hyperref}
\usepackage[thinlines]{easytable}

\newcolumntype{P}[1]{>{\centering\arraybackslash}p{#1}}

\def\V{\,\textrm{V}}
\def\eV{\,\textrm{eV}}
\def\meV{\,\textrm{meV}}
\def\Ry{\,\textrm{Ry}}

\newcommand{\rom}[1]{\uppercase\expandafter{\romannumeral #1\relax}}

\begin{document}

\title{Clar's goblet on graphene: field modulated charge transfer in a hydrocarbon heterostructure}
\author{Adam V. Bruce}
\author{Shuanglong Liu}
\author{James N. Fry}
\author{Hai-Ping Cheng}
\email{hping@ufl.edu}

\affiliation{Department of Physics, University of Florida, Gainesville, Florida 32611, USA}
\date{\today}

\begin{abstract}
In certain configurations, the aromatic properties of benzene ring structured molecules allow for unpaired, reactive valence electrons (known as radicals). Clar's goblets are such molecules. With an even number of unpaired radicals, these nanographenes are topologically frustrated hydrocarbons in which pi-bonding network and topology of edges give rise to the magnetism. Clar's goblets are therefore valued as prospective qubits provided they can be modulated between magnetic states. Using first principles DFT, we demonstrate the effects of adsorption on both molecule and substrate in a graphene-Clar's goblet heterostructure. We look at the energy difference bewteen FM and AFM states of the system and discuss underlying physical and chemical mechanisms in reference to the highest occupied molecular orbital (HOMO) and second HOMO (HOMO-1). We find that the HOMO of the molecule in the FM state is right at the Fermi surface, which leads to the hybridization between molecular state and the graphene state near the Dirac point. Furthermore, we investigate qualitative changes in charge realignment and magnetic state under variable electric field. Transitions from FM to AFM and back to FM states are observed.

\end{abstract}
\maketitle

\section{Introduction}
\label{sec:intro}

The abundant physicochemical properties of polycyclic aromatic hydrocarbons (PAH) have led to studies of these nanographenes (NGs) involving on-surface interaction.
Some PAHs have radical electrons, potentializing spintronic and quantum information science applications.
The choice of substrate is relevant, as interactions between molecule and substrate govern charge and spin transfer.
Additionally, given the carcinogenic nature of other hydrocarbons, strong adsorption of these molecules maintains environmental significance.
Some NG-substrate heterostructures experience little perturbation of molecular orbitals, as interaction of pi-bonds drives the physisorption process.
This phenomenon is well pronounced in the case of hydrocarbon adsorption on graphene and carbon nanotubes~\cite{Zhao2003}\textsuperscript{,}~\cite{Tournus2005}\textsuperscript{,}~\cite{Pei2013}\textsuperscript{,}~\cite{Lazar2013}\textsuperscript{,}~\cite{Wang2014}. First-principles studies of other ring-shaped organic molecules have produced similar results, showing the effect of geometry on adsorption~\cite{Woods2007}\textsuperscript{,}~\cite{Shtogun2007}.
Care should be taken to factor in the effects of symmetry breaking via the adsorption process; chirality can emerge from individually mirror-symmetric molecules and substrates~\cite{Richardson2007}.

The effects of adsorption on graphene are of interest, as charge doping from physisorption has been shown to reduce resistivity in single-walled carbon nanotubes~\cite{Zhao2003}. Graphene has been valued as a novel thin film semiconductor due to its special electronic properties resulting from the Dirac cone. Shifting of the Dirac cone through charge transfer has been observed in the adsoption of magnetic molecules on monolayer graphene\cite{Berkley2020}\textsuperscript{,}~\cite{Trinastic2014}. Adsorption of certain adatom dimers induces spin-dependence in graphene, leading to half-metalicity in the case of iron~\cite{Cao2010}. Adding ligands to metal molecules allows for varying levels of hole doping, shifting the cone above the Fermi energy~\cite{Trinastic2014, Scheerder2018, Li2014, Liu2020, Brooks2021}. The adsorption geometry is of note in the context of charge transfer, as seen in the adsorption of azobenzene; an external gating field induces differing charge doping between the isomers of this photoswitching molecule and graphene~\cite{Trinastic2014}. Gold clusters display a similar correlation, with \ce{Au_6} (commensurate packing with the graphene substrate) being physisorbed and yielding little charge transfer as opposed to \ce{Au_3} (vertical adsorption)~\cite{Scheerder2018}.

In the case of bilayer graphene, interlayer interaction breaks the degeneracy in the Dirac cone~\cite{McCann2006}. Experiments showing a full band gap opening under external gating field demonstrate the band structure programmability of this 2D structure~\cite{Zhang2009}\textsuperscript{,}~\cite{Mak2009}. At zero field, adsorption has been shown to induce a band gap on a moir\'{e} patterned monolayer~\cite{Balog2010}. While pristine graphene is agnostic toward spin polarization, edge defects can leave spins unpaired. Theoretical studies on bilayer graphene nanoribbons have demonstrated the effect of width on these edge states~\cite{Castro2008}\textsuperscript{,}~\cite{Xu2009}. Notably, the magnetic moment of edge states is dependent on ribbon width~\cite{Sahu2008}.

As with the magnetism in graphene nanoribbons, PAHs can exhibit magnetism induced by unbound electrons. One such PAH, \ce{C_38H_18}, or Clar's goblet, does not follow Kekul\'e structure but has two unpaired radicals\cite{Clar1972}. The non-zero spin makes the Clar's goblet a candidate for molecular spin qubit. The molecule contains 11 benzene rings in a goblet or bowtie shape, with radicals localized on each of the two benzo[cd]pyrene (BP) moieties\cite{Pogodin2003} and no aromaticity in the central ring. This localization allows for an anti-ferromagnetic (AFM) ground state, following non-Hund selection rules\cite{Ortiz2019}. Though larger PAHs can exhibit this topological frustration, Clar's goblet is among the smallest theoretically (via first-principles modeling) feasible to do so. Recent attempts to synthesize this nanographene have been successful, nearly half a century after the initial effort\cite{Mishra2020}.

We simulated the on-surface interaction of Clar's goblet on a pristine graphene substrate. Using first principles methods, we probed charge transfer at zero and finite electric fields. Structural relaxation was initiated from both stacking configurations seen in bilayer graphene, AA and AB types. Special attention was paid to perturbations of the molecule's radical orbitals. We employ arguments drawn from band structure, density of states, and real space charge analysis to demonstrate the physics of this heterostructure. Section II details the computational methods used throughout the paper. Section III consists of two subsections containing our computational results: III(A) probing adsorption of the molecule absent an external field and III(B) investigating the effect of electric field on adsorption. The final Section IV recounts the major results of the paper and discusses physical implications.

\section{Computational Methods}
\label{sec:method}

\begin{figure}[!b]
\centering
%\hspace*{-0.5 cm}
\includegraphics[width=8.5cm,height=\textheight, keepaspectratio]{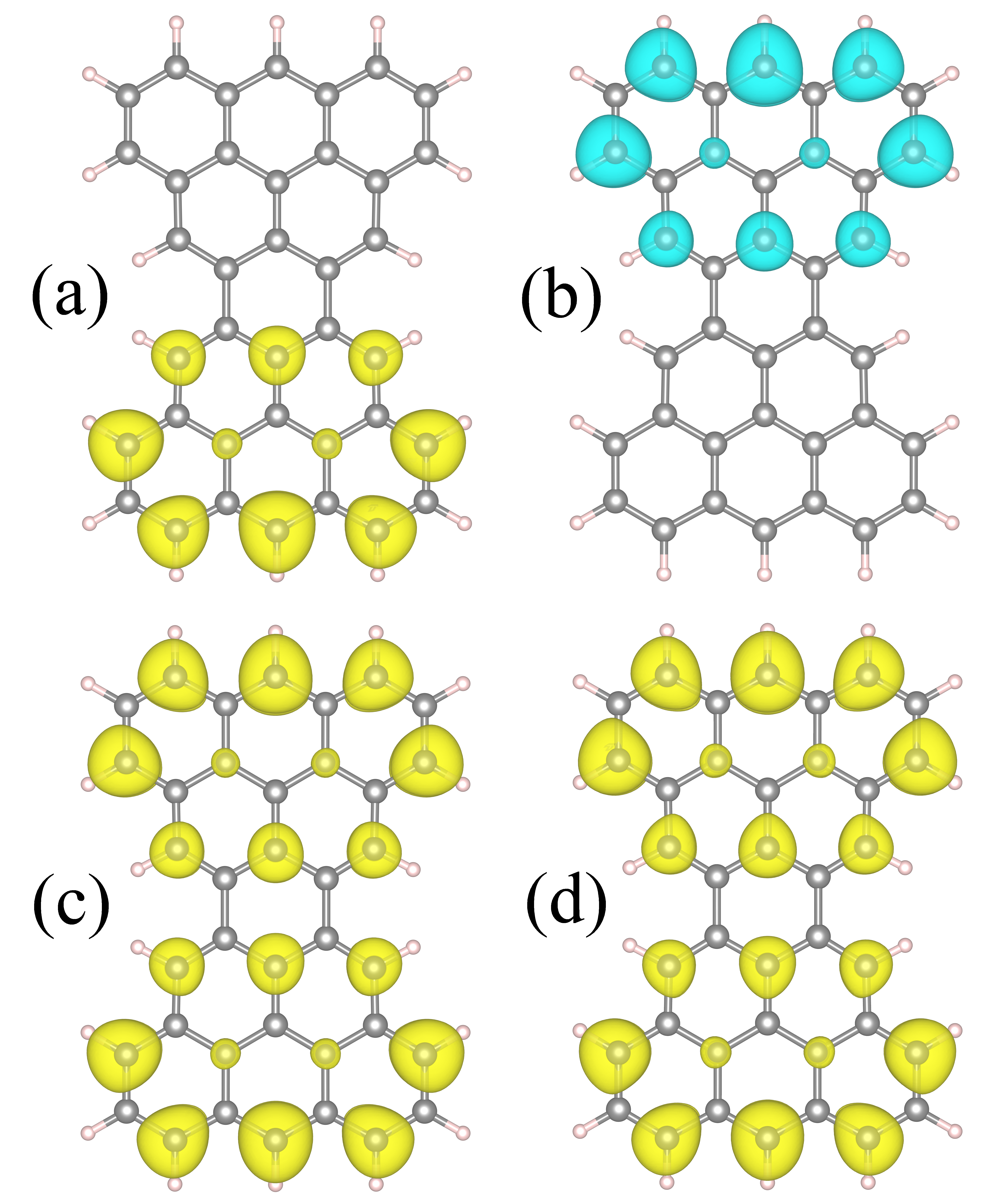}
\caption{\label{Figure orbitals} The radical orbitals' contribution to charge density of \ce{C_38H_18} Clar's Goblet; (a) up spin HOMO and (b) down spin HOMO-1 of the AFM state; (c) HOMO-1 and (d) HOMO of the FM state. All isosurfaces were generated in VESTA with an isosurface level of $5 \times 10\textsuperscript{-4}\,e$/\r{A}\textsuperscript{3}. Yellow isosurface designates a spin up electron, and teal a spin down.}
\end{figure}

Our calculations were performed on the first principles foundation of density functional theory (DFT)\cite{Kohn1965}. All physical quantities were derived from the Kohn-Sham Hamiltonian, which we obtained via self-consistent calculations in Quantum Espresso's (QE) PWscf package\cite{QE2009}\textsuperscript{,}\cite{QE2017}. The convergence threshold for total energy was set to $10^{-6}$ Ry between SCF steps. Atomic relaxation was also performed in QE using a $\Gamma$-point calculation with force and energy thresholds of $10^{-4} \Ry/a_0$ and $10^{-5} \Ry$ respectively. Kinetic energy cutoffs were $50 \Ry$ for wavefunctions and $400 \Ry$ for the charge density. The energy cutoffs were tested against the energy difference between magnetic states for values up to $160 \Ry$ for wavefunctions and $800 \Ry$ for density. Such an increase in precision and computational resources only increased the accuracy in energy difference by $10^{-5} \eV$. We employed the projector augmented wave (PAW) method and the Perdew-Burke-Ernzerhof (PBE) exchange correlation functional\cite{Blochl1994}\textsuperscript{,}\cite{PBE1996}. Our pseudopotentials \verb|C.pbe-n-kjpaw_psl.1.0.0.UPF| and \verb|H.pbe-kjpaw_psl.0.1.UPF| were taken from the QE website http://www.quantum-espresso.org. Calculations involving an applied electric field used the effective screening medium (ESM)\cite{Otani2006} with charge neutrality imposed globally.

\begin{figure}[!b]
\centering
%\hspace*{-0.5 cm}
\includegraphics[width=8.5cm,height=\textheight, keepaspectratio]{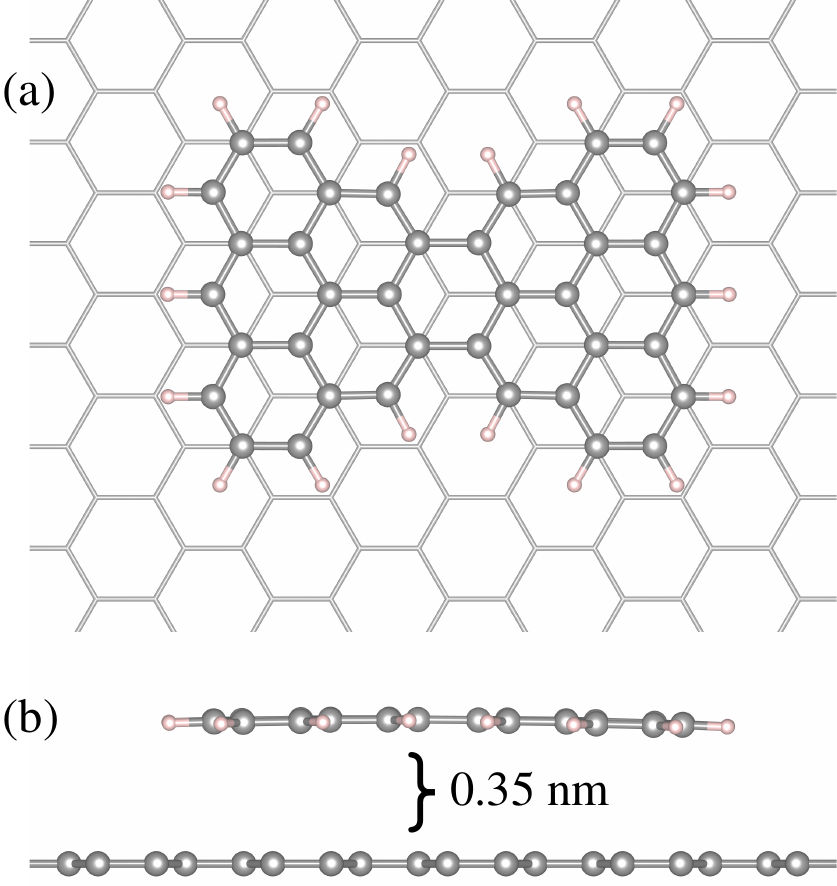}
\caption{\label{Figure AB_xyz} Heterostructure formed by adsorption of Clar's goblet on graphene, with the carbon atoms of the molecule presented in grey for contrast; (a) in-plane axis; (b) normal to plane axis; average equilibrium distance between materials is approximately 3.5 \r{A}.
The ground state AB-stacking introduces an asymmetry into the molecule}
\end{figure}

Our supercell was constructed with a $9 \times 12 \times 1$ cell of graphene and one \ce{C_38H_18} molecule. To capture the electron density of the full Brillouin zone, we used an adaptive $k$-mesh with greater quantity near the $\Gamma$ point at which the Dirac point of graphene is located.
In total, 39 irreducible k-points were employed in self-consistent calculations. This $k$-grid, however, was insufficient for visualization of density of states (DOS). We used a $24 \times 18 \times 1$ Monkhorst grid~\cite{Monkhorst1976} for the whole Brillouin zone and a 242-point adaptive mesh to capture the radical orbitals and Dirac cone, the reasoning for which will be explained in the Results section. Band structure, density of states, and real-space charge density were obtained with the QE post processing utilities pp.x, projwfc.x, and bands.x.

We computed charge transfer between the molecule and substrate using the Bader charge difference method that assigns charge to an atomic site within a surface minimizing electrostatic potential\cite{Bader1965}\textsuperscript{,}\cite{Bader1967}. We used the Bader code provided by the Henkelman group~\cite{Henkelman2006} to calculate this charge transfer from the electron density obtained through PWscf. As all of our calculations enabled spin polarization, we also investigated the possibility of spin transfer in the heterostructure.

\begin{figure}[b!]
\centering
%\hspace*{-0.5 cm}
\includegraphics[width=8.5cm,height=\textheight, keepaspectratio]{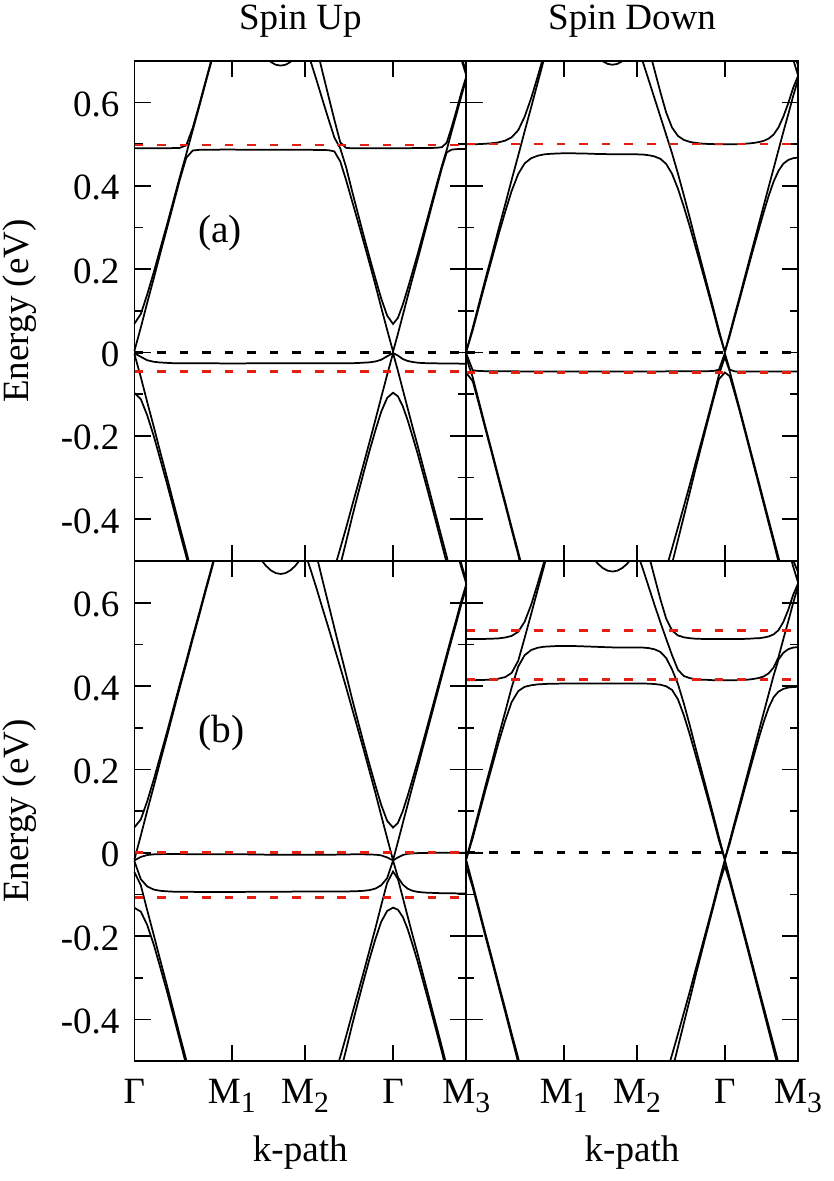}
\caption{\label{Figure AB_Bands} The band structure of the heterostructure in the (a) AFM and (b) FM states displaying spin up and -down channels. Molecular orbitals of the isolated molecule correspond to the red, horizontal bands. In units of the reciprocal lattice vectors $\boldsymbol{b}_1$ and $\boldsymbol{b}_2$, the special points $\textrm{M}_1$, $\textrm{M}_2$, and $\textrm{M}_3$ occur at $(0.5,0)$, $(0.5, 0.5)$, and $(0, 0.5)$ in the first Brillouin zone.
}
\end{figure}

\section{Results}
\label{sec:results}

\subsection{Zero Electric Field}
\label{sec:zerofield}

\begin{figure}[b!]
\centering
%\hspace*{-0.5 cm}
\includegraphics[width=8.5cm,height=\textheight, keepaspectratio]{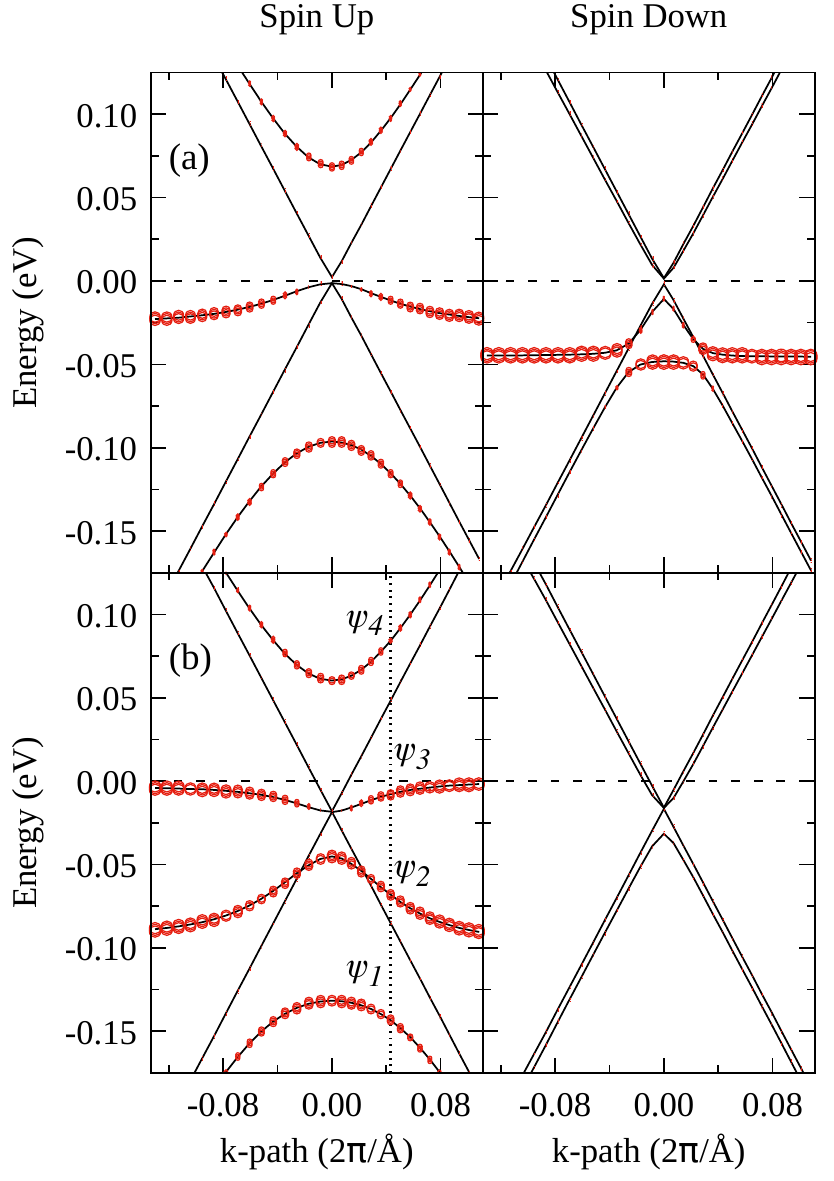}
\caption{\label{Figure AB_Fatbands} The $k$-resolved PDOS, also known as ``fatbands,'' of Clar's goblet juxtaposed on band structure. Panels (a) and (b) show the molecule fatbands of the AFM and FM state heterostructure respectively. The k-path is zeroed on $\Gamma$ with $M_2$ and $M_3$ along the negative and positive axes respectively. A dotted, vertical line in (b) marks the k-point $4.3 \times 10\textsuperscript{-2} \, 2\pi$\r{A}$^{-1}$ at which several bands show strong hybridization between graphene states and molecular orbitals.
}
\end{figure}

We simulated the adsorption of the \ce{C_38H_18} molecule Clar's goblet onto a monolayer graphene substrate. We first obtained the structure of the molecule relaxed in a vacuum. The carbon atoms were cut from graphene, and hydrogen was used to replaced the dangling bonds. Two initial magnetic states were prepared in order to obtain the known AFM and FM states. After vacuum relaxation, the molecule was placed on a graphene surface and subject to a second relaxation (with graphene atoms fixed). A final SCF calculation was performed on the isolated molecule, now with atomic positions taken from the molecule-substrate relaxation calculation. Analyzing partial charge density, we see in Fig.~\ref{Figure orbitals} the contribution of the HOMO and HOMO-1 to electron density in both magnetic states. Our results match well with previous DFT calculations on the molecule in a vacuum~\cite{Pogodin2003}, suggesting minimal perturbation of the radical orbitals. The energy of magnetic exchange ($E_\textrm{FM} - E_\textrm{AFM}$) was found to be $20.5 \meV$, an increase of $0.1 \meV$ from that of the vacuum-relaxed energy difference.

\
\onecolumngrid
\vskip 1\baselineskip

\
\begin{figure}[htbp!]
\centering
%\hspace*{-0.5 cm}
\includegraphics[width=0.85\textwidth,height=\textheight, keepaspectratio]{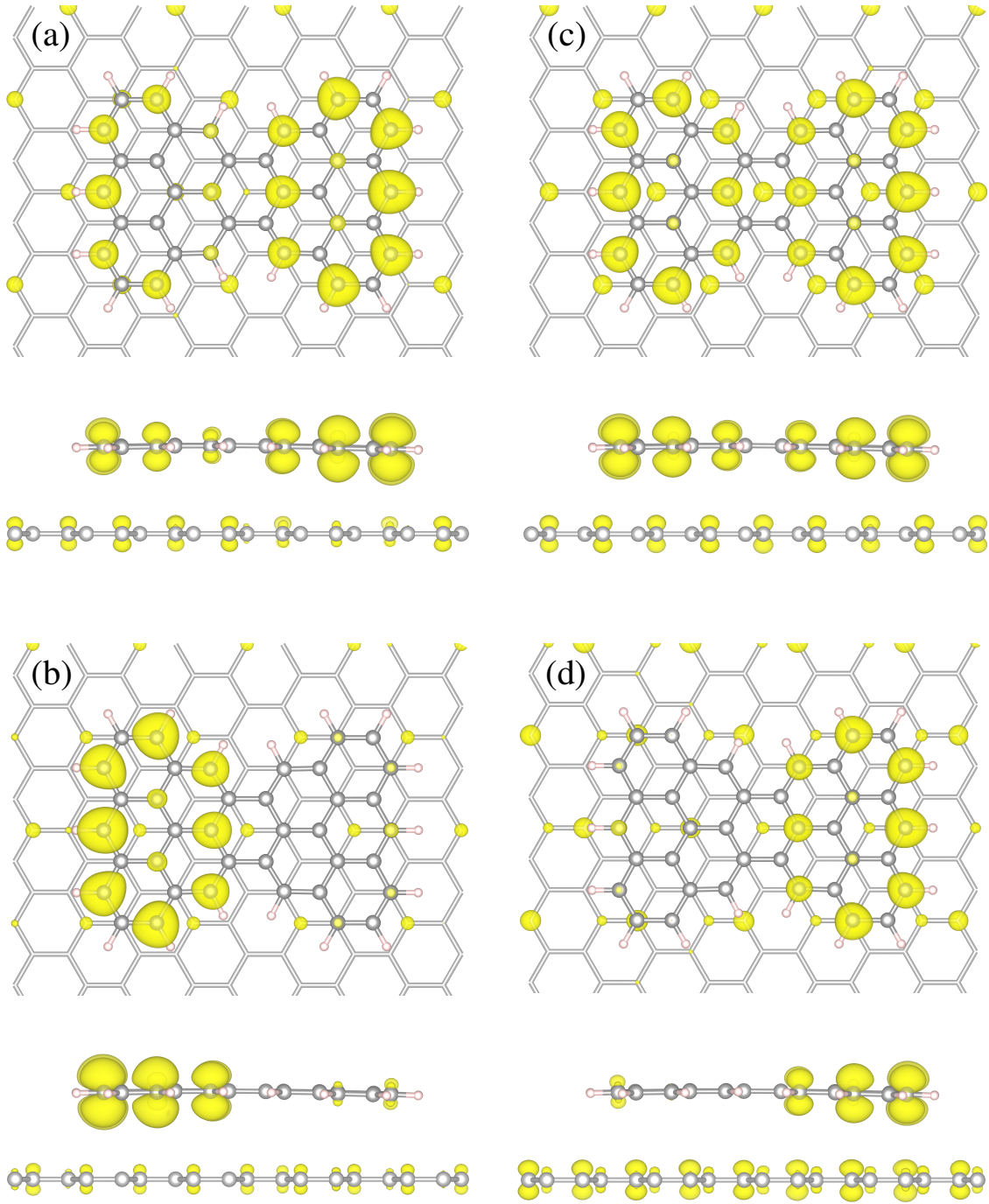}
\caption{\label{Figure WFC} (a)--(d) Partial charge density contribution from the Kohn-Sham states $\psi_1$, $\psi_2$, $\psi_3$, and $\psi_4$ respectively as marked in Fig.~\ref{Figure AB_Fatbands}b. The isosurface level is set to $5 \times 10\textsuperscript{-4}\,e$/\r{A}\textsuperscript{3} as noted in Fig.~\ref{Figure orbitals}.}
\end{figure}

\clearpage
\twocolumngrid

Our adsorption calculations probed the energetics of two commensurate stacking configurations based on AA and AB stacking. These two systems were prepared by aligning the carbon atoms of the molecule and graphene in the AA and AB positions of bilayer graphene. Both configurations are stable under structural relaxation, though the interlayer distance was reduced from the ansatz. Vertical adsorption geometry was not considered, as such bonding is expected to be much weaker than surface-to-surface interaction\cite{Wang2014}\textsuperscript{,}\cite{Leenaerts2009}. Like bilayer graphene, the AB stacking is the more energetically favorable configuration, with a total energy $158.7 \meV$ ($4.17 \meV$ per carbon atom of \ce{C_38H_18}) less than the AA stacking. The bonding energy of the molecule-substrate heterostructure is $1.966 \eV$. The preferred AB stacking configuration can be seen in Fig.~\ref{Figure AB_xyz}. We note some vertical deformation or displacement of 0.11 \r{A} of the molecule, which makes this stacking marginally different from true AB stacking. For simplicity, we will adhere to this nomenclature. The ground state remains anti-ferromagnetic with magnetic exchange energy $16.9 \meV$ in AB stacking. Previous studies show that radical localization on each BP moiety yields spin antisymmetry and symmetry in AFM and FM states respectively~\cite{Mishra2020}. The alternate stacking on each moiety breaks the D\textsubscript{2h} symmetry, allowing for a relative energy shift of the radical orbitals and the possibility of spin injection.

Fig.~\ref{Figure AB_Bands} shows the effects of adsorption on the band structure of the system. Both magnetic states show an opening of one Dirac cone in the spin up band structure, with a gap of approximately $0.2 \eV$ between open bands. This feature is characteristically similar to the $0.5 \eV$ band opening seen in bilayer graphene. It should be noted that the spin down population lacks this large band opening.
The molecular orbitals of Clar's goblet (red dashed lines) experience a minor shift in relative position. The AFM state (Fig.~\ref{Figure AB_Bands}a) sees an increase in energy disparity between HOMO and HOMO-1, while the FM state (Fig.~\ref{Figure AB_Bands}b) sees a decrease. Additionally, the relative shift in the Fermi energy indicates electron transfer from the molecule to graphene. 

\onecolumngrid
\vskip 1\baselineskip

\begin{figure}[b]
\centering
%\hspace*{-0.5 cm}
\includegraphics[width=17cm,height=\textheight, keepaspectratio]{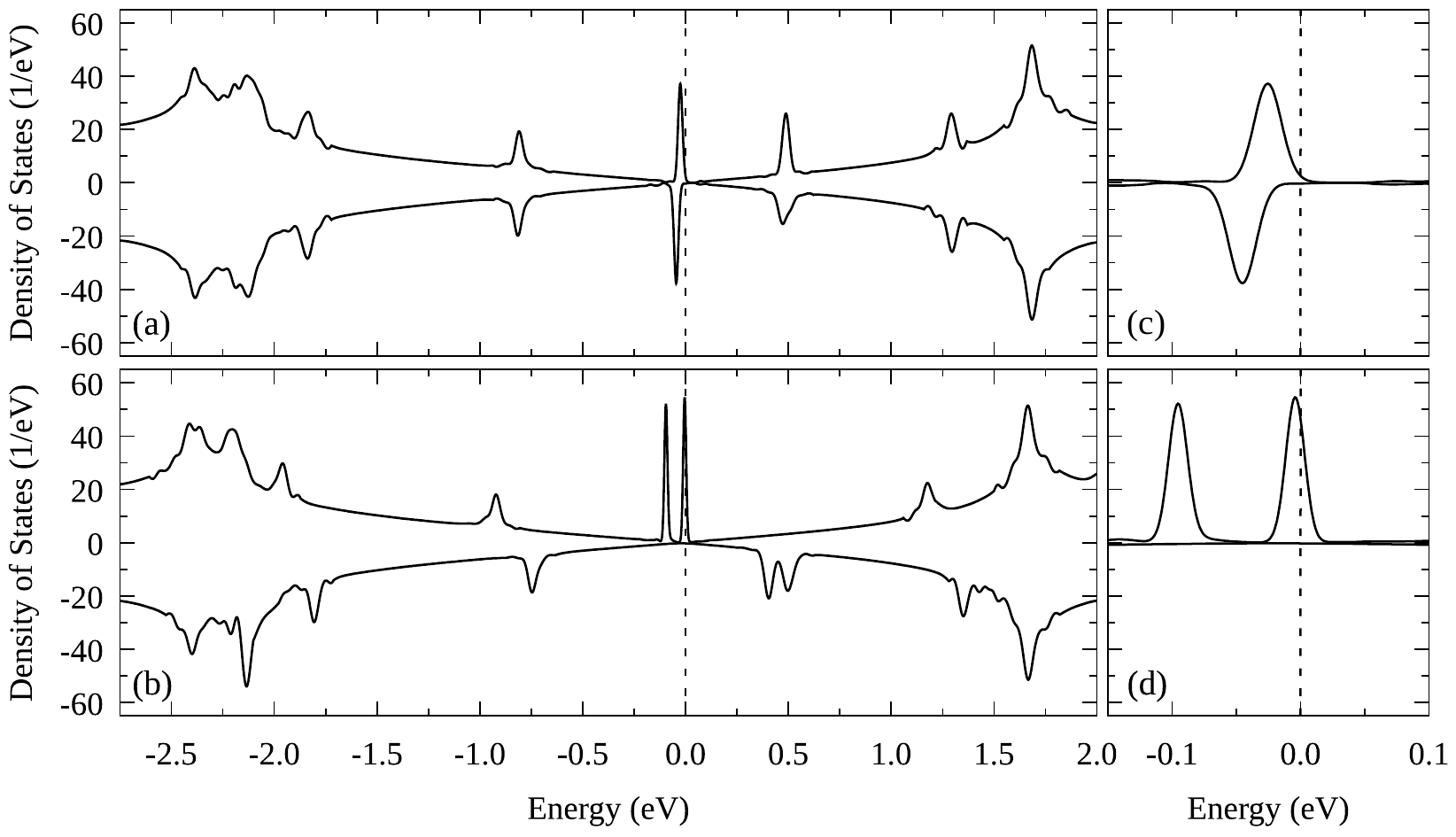}
\caption{\label{Figure AB_DOS} The density of states of the (a) AFM and (b) FM state heterostructure.
A variable smearing procedure is employed to illustrate the molecular orbital contribution to the  DOS.
Panels (c) and (d) show the DOS of (a) and (b) zoomed in near the Fermi energy.}
\end{figure}

\clearpage
\twocolumngrid

Differences in spin population are most pronounced at bands energetically adjacent to radicals. Fig.~\ref{Figure AB_Fatbands} presents the $k$-dependent partial density of states (PDOS) of the heterostructure derived from the projected DOS along the same $k$-path as the band structure (this collocation is often dubbed ``fatbands,'' as the point size of the bands is a function of the PDOS). Radical orbitals of the molecule can be identified by high PDOS beyond approximately $0.1 \, 2\pi$\r{A}$^{-1}$ in either direction of the k-path. States with little to no PDOS contribution from \ce{C_38H_18} (thin black lines) are mostly due to graphene. The molecule bands corresponding to the HOMOs experience dispersion and are strongly hybridized with graphene states near the $\Gamma$ point. In this region, the molecule yields greater PDOS than graphene along the broken Dirac cone states but substantially less along the ``HOMO band.'' This phenomenon is true for both magnetic states. Interestingly, although the AFM state radicals vary slightly in energy, the opening of the Dirac cone differs strongly between spins, with the AFM spin down HOMO-1 showing little hybridization with graphene bands. In FM state, the crossing of graphene bands corresponding to the Dirac cone occurs approximately at $-20 \meV$ rather than at the Fermi energy.

\begin{figure}[!b]
\centering
%\hspace*{-0.5 cm}
\includegraphics[width=8.5cm, keepaspectratio]{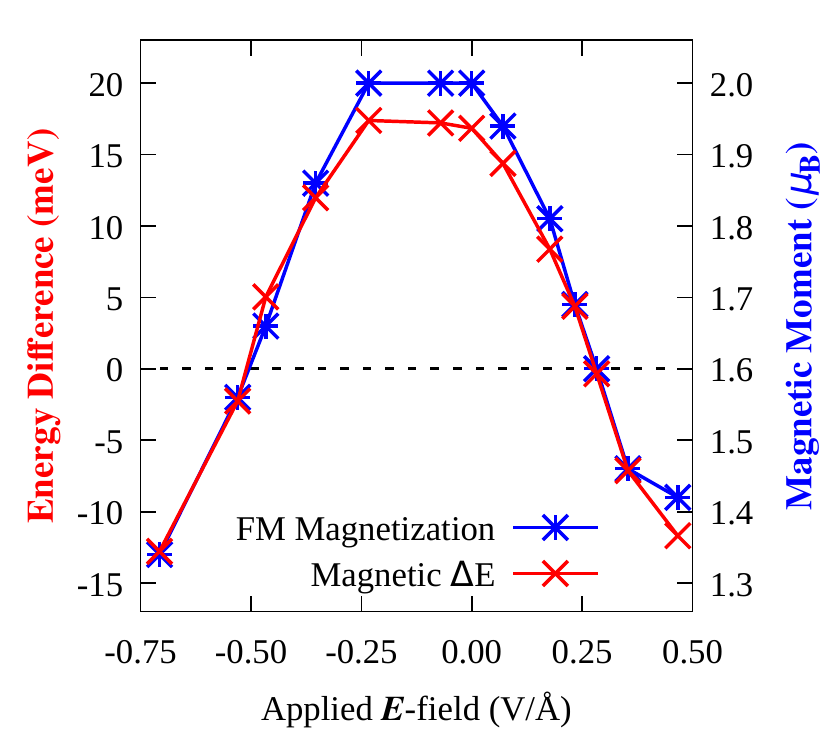}
\caption{\label{Figure ESM} Effects of applied electric field along the out-of-plane axis on the heterostructure AFM and FM states.
The FM state global magnetic moment per cell (blue asterisks) and energy difference between magnetic states, $\Delta E = E_\textrm{FM} - E_\textrm{AFM}$, (red crosses) are plotted against the strength and direction of $\bm{E}$-field. 
Negative (positive) magnetic exchange energy corresponds to a FM (AFM) ground state; the ground state becomes ferromagnetic around 0.28 and $-0.50 \V$/\r{A}. 
}
\end{figure}

In order to understand the nature of the FM state PDOS, we turned to charge density per Kohn-Sham state near (but not exactly at) the $\Gamma$ point. The state of Fig.~\ref{Figure WFC}a shows strong but unequal charge distribution on both BP moieties of Clar's goblet. Comparing with Fig.~\ref{Figure orbitals}c, this state displays charge on the same atomic sites as the molecular triplet state; however, the band corresponds to the broken Dirac cone. In Fig.~\ref{Figure WFC}b, the partial charge density along the molecule HOMO-1 band more closely resembles a radical of the AFM state Fig.~\ref{Figure orbitals}a. This stands in contrast with Fig.~\ref{Figure WFC}c, in which the charge distribution over the molecule matches the triplet state. The state in Fig.~\ref{Figure WFC}d, given by the unoccupied broken Dirac cone band, has charge patterned on the opposite BP moiety as Fig.~\ref{Figure WFC}b. Together, these four Kohn-Sham states contain features of the AFM and FM state radical orbitals, with $\psi_2$ and $\psi_4$ matching the former and  $\psi_1$ and $\psi_3$ the latter.

\def\tall{{}^{\strut}_{\strut}}

\begin{table}[!b]
%\vskip 1\baselineskip
%\begin{tabular}{ | m{0.85cm} | m{1cm}| m{1.25cm} | m{1.6cm} | m{1.4cm} | m{1.6cm} | }
\begin{tabular}{|m{0.835cm}|m{0.985cm}|m{3.06cm}|m{3.06cm}|}
\hline
State & $\bm{E}$-field &\ce{C_38H_18} Clar's Goblet & Graphene Substrate$\tall$ \\
\hline
\end{tabular}
\begin{TAB}(r,1.1cm,1.1cm)[3.2pt]{|c|c|c|c|c|c|}{|c|c|c|c|c|c|c|c|c|}
  %\hline
  %State & $\bm{E}$-field (V/\r{A}) &\ce{C_38H_18} Spin Up &\ce{C_38H_18} Spin Down & Graphene Spin Up & Graphene Spin Down \\
  %State & $\bm{E}$-field &\ce{C_38H_18} &\ce{C_38H_18} & Graphene & Graphene \\
  %\hline
   & (V/\r{A}) & Spin Up & Spin Down & Spin Up &  Spin Down \\
  FM & $-0.70$ & $+0.00e$ & $+0.75e$ & $-0.33e$ & $-0.42e$\\ %\tall$ \\
  %\hline
  FM & $-0.35$ & $-0.01e$ & $+0.20e$ & $-0.06e$ & $-0.13e$\\ %\tall$ \\
  %\hline
  FM & $+0.00$ & $-0.06e$ & $+0.01e$ & $+0.06e$ & $-0.01e$\\ %\tall$ \\
  %\hline
  FM & $+0.35$ & $-0.57e$ &$+0.00e$ & $+0.30e$ & $+0.27e$\\ %\tall$ \\
  %\hline
  AFM & $-0.70$ & $+0.39e$ & $+0.34e$ & $-0.32e$ & $-0.41e$\\ %\tall$ \\
  %\hline
  AFM & $-0.35$ & $+0.09e$ & $+0.08e$ & $-0.07e$ & $-0.10e$\\ %\tall$ \\
  %\hline
  AFM & $+0.00$ & $-0.04e$ & $+0.00e$ & $+0.04e$ & $+0.00e$\\ %\tall$ \\
  %\hline
  AFM & $+0.35$ & $-0.30e$ & $-0.25e$ & $+0.28e$ & $+0.27e$\\ %\tall$ \\
  %\hline
\end{TAB}
%\end{tabular}
\caption{\label{tab_esm} Bader charge transfer between \ce{C_38H_18} Clar's goblet and graphene substrate under variable electric field; values shown depict doping per unit cell of the heterostructure.
The zero-field data reflects charge reordering from adsorption alone}
\end{table}

The energy shift in band structure occurs within a small energy range, necessitating a resolution finer than $20 \meV$.  First principles DOS calculations involving graphene require a relatively dense $k$-mesh to capture the Dirac cone; sparse sampling of $k$-points erroneously yields occupation at the Fermi energy, but the computational cost of a sufficiently dense uniform sampling over the entire Brillouin zone can far exceed that of SCF calculations, particularly in the case of graphene heterostructures. We achieved high resolution within an energy window encompassing the radical orbitals via an adaptive $k$-mesh. The densest sampling was positioned about the $\Gamma$ point, capturing all the graphene bands within the energy window. The molecule bands are mostly flat outside this range (as previously shown in Fig.~\ref{Figure AB_Bands}), requiring very few $k$-points. In order to account for the full Brillouin zone, we performed a second DOS calculation with uniform $k$-mesh. Fig.~\ref{Figure AB_DOS} shows the result, in which the two regimes are subject to different smearing parameters appropriate to the energy resolution. The density of states supports the band structure result of HOMO positioning with respect to the Dirac cone. The HOMO of the FM state (Fig.~\ref{Figure AB_DOS}a) is partially occupied, peaking around $-4 \meV$. In the AFM case (Fig.~\ref{Figure AB_DOS}b), the radicals energetically vary by approximately $20 \meV$, with HOMO centered around $-25 \meV$. Several other peaks corresponding to molecular orbitals can be distinguished from the graphene DOS, differing cosmetically from the radicals due to a discrepancy in smearing. The intact pi-orbitals and relatively low charge transfer indicate a physisorption mechanism.

\onecolumngrid
\vskip 1\baselineskip

\begin{figure}[b]
\centering
%\hspace*{-0.5 cm}
\includegraphics[width=0.9\textwidth,height=\textheight, keepaspectratio]{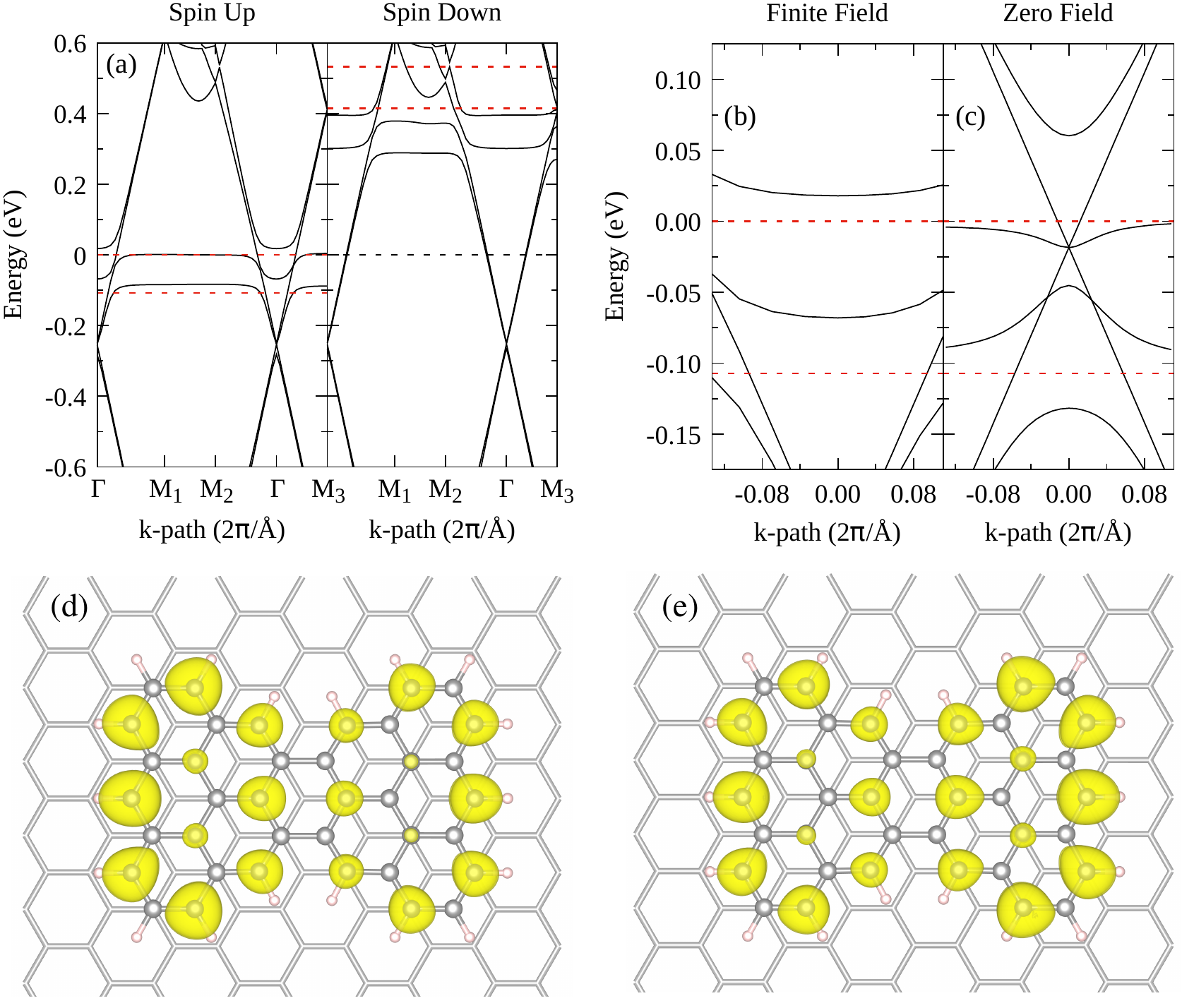}
\caption{\label{Figure ESM_bands} (a) The band structure of the FM ground state at $0.35 \V$/\r{A}; (b)--(c) comparison of the spin up band structure around the $\Gamma$ point with the zero-field case; spin up partial charge density for (d) the highest occupied Kohn-Sham state at $\Gamma$ and for (e) the lowest unoccupied Kohn-Sham wavefunction at $\Gamma$; bands at the tip of the Dirac cone are wholly localized to graphene}
\end{figure}

\clearpage
\twocolumngrid

\subsection{Finite Electric Field}
\label{sec:finitefield}

Using the ESM method in Quantum Espresso, we introduced an electric field normal to the graphene plane, with a positive field directed toward the \ce{C_38H_18} molecule. SCF calculations initialized the magnetic states to FM and AFM but allowed for changes in magnetization. Sampling the energies of both AA and AB stacking at strong applied fields, we found that the ground state stacking configuration remains AB; all subsequent simulations used the AB commensurate stacking discussed in III(A). Fig.~\ref{Figure ESM} tracks the exchange coupling energy and magnetic moment of the ferromagnetic state at field strengths up to $0.7 \V$/\r{A}. The energy disparity between magnetic states is relatively unchanged on the interval from $0$ to $-0.25 \V$/\r{A} (electric field directed from molecule to substrate). Beyond $-0.25 \V$/\r{A} the exchange energy is steadily reduced with increasing electric field until the state becomes ferromagnetic $-0.50 \V$/\r{A}. In the opposite direction, the field induces charge transfer beginning at relatively small field strengths. A change in the magnetic ground state at occurs at $0.28 \V$/\r{A}. The incident field reduces the global magnetic moment of the FM state regardless of field direction beyond $\pm 0.25 \V$/\r{A}, and the change in ground state correlates with magnetization $1.6 \mu_B$ per unit cell. We understand the asymmetry in charge transfer and field direction to be a consequence of the HOMO-LUMO gap seen in Fig.~\ref{Figure AB_DOS}. The HOMO is pinned to the Fermi energy, but the LUMO occurs much higher above the Fermi energy. A substantial change in DOS is needed to allow electron doping of the molecule. Consequently, the energy difference in Fig.~\ref{Figure ESM} plateaus between 0 to $-0.25 \V$/\r{A}.

As the ground state becomes ferromagnetic for both field orientations, we conclude that the FM state is generally more susceptible to charge realignment under gating field than the AFM state. Table~\ref{tab_esm} breaks down the charge transfer between molecule and substrate as well as spin population using Bader analysis. The total charge doping on graphene per atom (with 216 atoms in the graphene supercell) is comparatively small under the $0.0 \V$/\r{A} field, and the substrate resists spin injection. In the AFM state, the molecule experiences spin transfer involving both up and down populations. Spin transfer in the FM state follows a Hund-like convention, filling the down spin population under electron doping and losing up spin under hole doping. The global magnetic moment reported in Fig.~\ref{Figure ESM} correlates with charge transfer on Clar's goblet. Extrapolating the magnetic moment from spin transfer, the magnetization of the molecule differs from global magnetization by approximately 0.1 Bohr magneton per unit cell of the heterostructure. 

Effects of the electric field on the band structure of the FM state can be seen in Fig.~\ref{Figure ESM_bands}a. The Dirac cone of graphene is intact and electron doped. At the $\Gamma$ point, the highest occupied and lowest unoccupied Kohn-Sham states (along the bands of Fig.~\ref{Figure ESM_bands}b) form an energy gap comparable to the HOMO-1 and HOMO (the red, dashed lines). Fig.~\ref{Figure ESM_bands}d--e shows the contribution to charge density from these states. Though charge on each BP moiety is uneven, these states are recognizable as modifications of the FM radicals. Further, the charge density from these states is localized to the molecule. The state in Fig.~\ref{Figure ESM_bands}e can be identified as both the HOMO of the molecule and charge donor to the graphene states. Changes in spin up band structure away from the $\Gamma$ point are less noticeable; a smaller energy difference separates the bands of radical orbitals along the edge of the Brillouin zone. For spin down, Fig.~\ref{Figure ESM_bands}a shows the LUMO and LUMO+1 shifted closer to the Fermi energy by more than $0.1 \eV$.

\section{Conclusions}
\label{sec:conclusions}

We simulated a heterostructure composed of \ce{C_38H_18} Clar's goblet and a graphene monolayer supercell. AB-type stacking was determined to be the ground state stacking configuration, and the magnetic ground state of the molecule remained antiferromagnetic. Adsorption under zero field broke the degeneracy in graphene's Dirac cones only for spin up population. The HOMO was found pinned to the Fermi energy and showed hybridization with the substrate Dirac cone states. Turning to partial charge density, the breaking of the Dirac cone occurred due to hybridization of graphene bands with the radicals. The HOMO-1 experienced notable change in density, with charge distribution on each BP moiety altered near $\Gamma$. Bader charge analysis showed charge transfer of approximately 0.05 electrons from molecule to graphene, which is in agreement with the shift in band structure. Given the binding energy of nearly $2 \eV$, the calculated charge transfer was too small to indicate ionic bonding. Density of states further reinforced this claim, revealing separability between Clar's goblet and graphene DOS. For the molecular orbitals, our results indicated preservation of relative energy levels along the intervals from $-2.0 \eV$ to the HOMO-1 as well as from the Fermi level to $1.5 \eV$. The radicals experience slight modification during adsorption, most notably increasing the energy disparity between HOMO-1 and HOMO in the AFM state. However, this energetic modification (order $20 \meV$) should be contrasted with the well-preserved HOMO-LUMO gap (order $0.4 \eV$).

Under external gating field, the graphene substrate served as a reservoir for charge doping without altering the fundamental physical and chemical properties of this nanographene. The direction of the field controlled the direction of charge transfer. Though both electron and hole doping were admitted, a greater field strength was required to electron dope Clar's goblet. Spin transfer is dependent on the energy levels of molecular orbitals as confirmed by Bader analysis. Band structure of the system under the finite field showed that electron doping does not induce a band opening in graphene states. The bands of the molecule and substrate were shifted by the field but retained their individual characteristics. Hybridization occurs only near the intersection in energy and k-path of molecule radicals and substrate bands.

Clar's goblet was shown to undergo a transition in magnetic ground state at sufficiently high field strength in either direction. A ferromagnetic state becomes energetically preferred for greater than $0.4 e$ charge transfer. In this configuration electron doping fills the spin down LUMO, and hole doping depletes the spin up HOMO. As a result, the FM state magnetic moment decreases for any direction of charge transfer. The AFM state, however, receives spin injection from both populations in roughly equal quantity. Both radicals of the AFM state become partially occupied (inferred from Bader analysis) while only the HOMO of the FM state donates charge (confirmed by band structure). Further investigation is necessary to identify the physical mechanism of the magnetic exchange.

\begin{acknowledgments}

This work was supported by the US Department of Energy (DOE), 
Office of Basic Energy Sciences (BES), under Contract No. DE-SC0022089. 
Computations were done using the utilities of National Energy Research Scientific Computing Center 
and University of Florida Research Computing.

\end{acknowledgments}

%\appendix
%\renewcommand{\thefigure}{\thesection\arabic{figure}}
%\renewcommand{\thetable}{\thesection\arabic{table}}

\bibliographystyle{apsrev4-2}
\bibliography{ref}

\end{document}